\documentstyle[sprocl,axodraw]{article}
\def\Journal#1#2#3#4{{#1} {\bf #2}, #3 (#4)}
\def\NPB{{\em Nucl. Phys.} B}

\def\PRD{{\em Phys. Rev.} D}

\def\ra{\rightarrow}

\def\beq{\begin{equation}}
\def\eeq{\end{equation}}
\def\barr{\begin{eqnarray}}
\def\earr{\end{eqnarray}}
\newcommand{\newc}{\newcommand}
\newc{\etal}{{\it et al.}\ }
\newc{\eg}{{\it e.g.}\ }
\newc{\ie}{{\it i.e.}\ }
\newc{\cf}{{\it cf.}\ }
\newc{\lam}{\lambda}
\newc{\eps}{\epsilon}
\newc{\gev}{\,GeV}
\newc{\Ebar}{{\bar E}}
\newc{\Dbar}{{\bar D}}
\newc{\Ubar}{{\bar U}}
\newc{\rp}{$R_p$}
\newc{\half}{\frac{1}{2}}
\newc{\rpv}{{\not \!\! R_p}}
\newc{\rpvm}{{\not  R_p}}
\newc{\eq}[1]{(\ref{eq:#1})}
\newc{\lab}[1]{\label{eq:#1}}
\newc{\gsim}{{\stackrel{>}{\sim}}}
\newc{\lsim}{{\stackrel{<}{\sim}}}
\newc{\nonum}{\nonumber}
\newc{\rpmssm}{{$R_p$-MSSM}\ }
\newc{\rpvmssm}{$\rpv$-MSSM}
\begin{document}
\title{An Introduction to Explicit R-parity Violation}
\author{ Herbi Dreiner} 
\address{Rutherford Appleton Laboratory, Chilton, Didcot, Oxon OX11 0QX, UK}
%%%%%%%%%%%%%%%%%%%%%%%%%%%%%%%%%%%%%%%%%%%%%%%%%%%%%%%%%%%%%%
% You may repeat \author \address as often as necessary      %
%%%%%%%%%%%%%%%%%%%%%%%%%%%%%%%%%%%%%%%%%%%%%%%%%%%%%%%%%%%%%%

\maketitle

\abstracts{
I discuss the theoretical motivations for R-parity violation, review the
experimental bounds and outline the main changes in collider phenomenology
compared to conserved R-parity. I briefly comment on the effects of R-parity
violation on cosmology. 
}

\section{Introduction}
Until recently, R-parity violation ($\rpv$) has been considered an unlikely
component of the supersymmetric extension of the Standard Model (SM). In the 
past two years, it has motivated potentially favoured solutions to 
experimentally observed discrepancies (\eg $R_b,\,R_c$, ALEPH four-jet events,
HERA high $Q^2$ excess). It is the purpose of this chapter to present $\rpv$ as
an equally well motivated supersymmetric extension of the SM and provide an
introductory guide. I start out with the definition of \rp\ and the most serious
problem of proton decay. Then I discuss the various motivations for $\rpv$,
contrasting them with the \rp-conserving MSSM. Afterwards, I give an overview of
the phenomenology of $\rpv$. I finish with a discussion on cosmological effects.

\section{What is R-parity?}
R-parity (\rp) is a discrete multiplicative symmetry. It can be written as
\cite{rp}
\beq
R_p= (-{\bf 1})^{3B+L+2S}.
\lab{rparity}
\eeq
Here $B$ denotes the baryon number, $L$ the lepton number and $S$ the spin of a
particle. The electron has $R_p=+1$ and the selectron has $R_p=-1$. In fact, for
all superfields of the supersymmetric SM, the SM field has $R_p=+1$ and its 
superpartner has \footnote{In general symmetries for which the anticommuting
parameters, $\theta$, transform non-trivially (and thus superpartners
differently) are denoted R-symmetries. They can be discrete (\rp), global 
continuous, or even gauged \cite{gaugedr,gaugedrus}. R-symmetries can be broken
without supersymmetry being broken.} $R_p=-1$. \rp\ is
conserved in the MSSM, superpartners can only be produced in pairs (all initial
states at colliders are \rp\  even) and the LSP is stable. When extending the
SM with supersymmetry one doubles the particle content to accomodate the
superpartners and adds an additional Higgs doublet superfield. The minimal
symmetries required to construct the Lagrangian are the gauge symmetry of the
SM: $G_{SM}=SU(3)_c\times SU(2)_L\times U(1)_Y$ and supersymmetry (including
Lorentz invariance). The most general superpotential with these symmetries and
this particle content (\cf Ch. 1) is \cite{superpot} 
\barr
W &=& W_{MSSM} + W_\rpvm,
\lab{superpot} \\
W_{MSSM}&=&h^e_{ij}L_iH_1\Ebar_j+ h^d_{ij}Q_iH_1\Dbar_j+ h^u_{ij}Q_iH_2\Ubar_j+
\mu H_1H_2, \lab{mssmsuper} \\
W_\rpvm &=& \half \lam_{ijk} L_iL_j\Ebar_k + \lam'_{ijk} L_iQ_j\Dbar_k +
\half \lam''_{ijk} \Ubar_i\Dbar_j\Dbar_k+\kappa_iL_iH_2.
\lab{rpvsuper1}
\earr
$i,j=1,2,3$ are generation indices and a summation is implied. $L_i$ ($Q_i$) are
the lepton (quark) $SU(2) _L$ doublet superfields. $\Ebar_j$ ($\Dbar_j,\Ubar_j$)
are the electron (down- and up-quark) $SU(2)_L$ singlet superfields. $\lam,\,
\lam',$ and $\lam''$ are Yukawa couplings. The $\kappa_i$ are dimensionful mass
parameters. The $SU(2)_L$ and $SU(3)_C$ indices have been suppressed. When
including them we see that the first term in $W_\rpvm$ is anti-symmetric in $\{
i,j\}$ and the third term is anti-symmetric in $\{j,k\}$. Therefore $i\not=j$ in
$L_iL_j\Ebar_k$ and $j\not= k$ in $\Ubar_i\Dbar_j\Dbar_k$. Eq.\eq{rpvsuper1}
thus contains $9+27+9+3=48$ new terms beyond those of the MSSM. 

The last term in Eq.\eq{rpvsuper1}, $L_iH_2$, mixes the lepton and the Higgs
superfields. In supersymmetry $L_i$ and $H_1$ have the same gauge and Lorentz
quantum numbers and we can redefine them by a rotation in $(H_1,L_i)$. The terms
$\kappa_iL_iH_2$ can then be rotated to zero in the superpotential 
\cite{hallsuzuki}. If the corresponding soft supersymmetry breaking parameters $
B_i$ are aligned with the $\kappa_i$ they are simultaneously rotated away
\cite{hallsuzuki,yuval}. However, the alignment of the superpotential terms with
the soft breaking terms is not stable under the renormalization group equations
\cite{white}. Assuming an alignment at the unification scale, the resulting
effects are small \cite{white} except for neutrino masses \cite{white,nilles}.
The effects can be further suppressed by a horizontal symmetry. Throughout the
rest of this Chapter, I will assume the $L_iH_2$ terms have been rotated away
\footnote{The ambiguity on bounds due to rotations in $(L_i,H_1 )$ space has
been discussed in \cite{davidsonellis}.} 
\beq
W_\rpvm= \lam_{ijk} L_iL_j\Ebar_k + \lam'_{ijk} L_iQ_j\Dbar_k +\lam''_{ijk}
\Ubar_i\Dbar_j\Dbar_k.
\lab{rpvsuper}
\eeq
Expanding for example the $LL\Ebar$ term into the Yukawa couplings yields
\beq
{\cal L}_{LL\Ebar}=\lam_{ijk}\left[ {\tilde\nu}^i_L{\bar e}^k_Re_L^j
+{\tilde e}_L^j{\bar e}_R^k\nu^i_L + ({\tilde e}_R^k)^* ({\bar\nu}_L^i)^ce_L^j
- (i\leftrightarrow j)\right]+h.c.
\lab{lleexp}
\eeq
The tilde denotes the scalar fermion superpartners. These terms thus violate 
lepton-number. The $LQ\Dbar$ terms also violate lepton number and the $\Ubar
\Dbar\Dbar$ terms violate baryon number. The entire superpotential 
\eq{rpvsuper} violates \rp. 

\section{Proton Decay and Discrete Symmetries}
The combination of lepton- and baryon-number violating operators in the
Lagrangian can possibly lead to rapid proton decay. For example the two 
operators $L_1Q_1\Dbar_k$ and $\Ubar_1\Dbar_1\Dbar_k$ ($k\not=1$) can 
contribute to proton decay via the interaction shown in Figure 1a. 
\begin{figure}[t]
\begin{center}
\begin{picture}(340,85)(0,0)
\SetWidth{1.5}
%
% Proton Decay
%
\ArrowLine(12,14)(40,56)
\ArrowLine(12,99)(40,56)
\ArrowLine(98,99)(70,56)
\ArrowLine(98,14)(70,56)
\DashLine(40,56)(70,56){3}
\Text(3,18)[]{{\Large $d$}}
\Text(3,94)[]{{\Large $u$}}
\Text(110,98)[]{{\Large $e^+$}}
\Text(108,21)[]{{\Large ${\bar u}$}}
\Text(57,65)[]{{\Large${\bar{\tilde s}}$ }}
\Text(57,0)[]{{\large(a)}}
%
% tau Decay
%
\ArrowLine(125,71)(155,71)
\ArrowLine(201,104)(155,71)
\DashLine(155,71)(175,42){3}
\ArrowLine(175,42)(203,60)
\ArrowLine(175,43)(203,14)
\Text(135,83)[]{{\Large $\tau^-$ }}
\Text(214,63)[]{{\Large $e^-$}}
\Text(214,109)[]{{\Large ${\bar\nu}_e$}}
\Text(214,18)[]{{\Large $\nu_\tau$}}
\Text(176,63)[]{{\Large${{\tilde e}_k}$ }}
\Text(175,0)[]{{\large(b)}}
%
% Neutralino Decay
%
\ArrowLine(235,71)(265,71)
\ArrowLine(265,71)(311,104)
\DashLine(265,71)(285,42){3}
\ArrowLine(285,42)(313,60)
\ArrowLine(313,14)(285,42)
\Text(250,83)[]{{\Large ${\tilde\chi}^0$}}
\Text(328,107)[]{{\Large $e^-$ }}
\Text(285,63)[]{{\Large ${\tilde e}$}}
\Text(325,62)[]{{\Large $u$}}
\Text(324,18)[]{{\Large ${\bar d}$}}
\Text(285,0)[]{{\large(c)}}
\end{picture}
\end{center}
\caption{(a) Proton decay via $\Ubar_1\Dbar_1\Dbar_2$ and $L_1Q_1\Dbar_2$, (b)
Tau decay via two $L_1L_3\Ebar _k$ insertions, (c) Neutralino decay via
$L_1Q_1\Dbar_1$.} 
\end{figure}
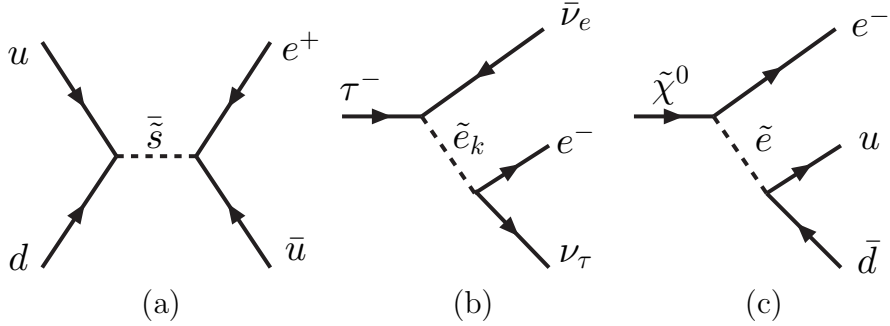
On  dimensional grounds we estimate 
\beq
\Gamma (P\ra e^+\pi^0)\approx\frac{\alpha(\lam'_{11k})\alpha(\lam''_{11k})}
{{\tilde m}_{dk}^4}{M_{proton}^5}.
\eeq
Here $\alpha(\lam)=\lam^2/(4\pi)$. Given that \cite{pdg} $\tau(P\ra e\pi)
>10^{32}\,yr$, we obtain 
\beq
\lam'_{11k}\cdot\lam''_{11k}\; \lsim\; 2\cdot10^{-27} \left( \frac{{\tilde
m}_{dk}} {100\gev} \right)^2. \lab{proton}
\eeq
For a more detailed calculation see \cite{sher}. This bound is so strict that
the only natural explanation is for at least one of the couplings to be zero.
Thus the simplest supersymmetric extension of the SM is excluded: an extra
symmetry is required to protect the proton. 

In the MSSM, \rp\ is imposed by hand. This forbids all
the terms in $W_\rpvm$ and thus protects the proton. An alternative discrete
 symmetry with the same physical result is matter parity
\beq
(L_i,\Ebar_i,Q_i,\Ubar_i,\Dbar_i) \ra - (L_i,\Ebar_i,Q_i,\Ubar_i,\Dbar_i), 
\quad (H_1,H_2)\ra (H_1,H_2).
\eeq
This forbids all terms with an odd power of matter fields and thus forbids all
the terms in $W_\rpvm$. However, there are other solutions, which protect the
proton equally well. If baryon number is conserved the proton can not decay.
Thus forbidding just the interactions $\Ubar_i\Dbar_j\Dbar_k$ is sufficient.
This can be achieved by baryon-parity 
\beq
(Q_i,\Ubar_i,\Dbar_i) \ra - (Q_i,\Ubar_i,\Dbar_i),\quad(L_i,\Ebar_i,H_1,H_2)
\ra (L_i,\Ebar_i,H_1,H_2).
\lab{bp}
\eeq
This symmetry thus protects the proton but allows for $\rpv$ via the $L_iL_j
\Ebar_k$ and $L_iQ_j\Dbar_k$ operators. If only the the interactions 
$\Ubar_i\Dbar_j\Dbar_k$ are allowed {\it and} the proton is lighter than the 
LSP the proton is stable as well. This can be achieved by lepton parity 
\beq
(L_i,\Ebar_i) \ra - (L_i,\Ebar_i), \quad
(Q_i,\Ubar_i,\Dbar_i,H_1,H_2) \ra (Q_i,\Ubar_i,\Dbar_i,H_1,H_2).
\lab{lp}
\eeq
Baryon-parity and lepton parity are two possible solutions to maintain a stable
proton {\it and} allow for $\rpv$. There is a large number of discrete
symmetries which can achieve this \cite{vissani}.

\section{Motivation}
The symmetries discussed in the previous section were all imposed {\it ad hoc}
with no deeper motivation than to ensure the stability of the proton. On this
purely phenomenological level there is no reason to prefer the models with
conserved \rp\  versus those with $\rpv$. However, this is not a satisfactory
view of the weak-scale picture. Hopefully, the correct structure will emerge
{}from a simpler theory at a higher energy predicting either \rp-conservation or
$\rpv$. 

{\it Grand Unified Theories:}\ \ 
In GUTs quarks and leptons are typically in common multiplets and thus have the
same quantum numbers. The discrete symmetries protecting the proton and
resulting in $\rpv$ typically treat quarks and leptons differently and thus seem
incompatible with a GUT. All the same, several GUT models have been constructed
\cite{hallsuzuki,brahmhall,rattazzi,rest} which have low-energy $\rpv$. This is
typically achieved by non-renormalisable GUT scale operators involving Higgs
fields. These operators become renormalisable $\rpv$-operators after the GUT
symmetry has been broken. Such models have been constructed for the GUT gauge
groups $SU(5)$ \cite{hallsuzuki,rattazzi}, $SO(10)$ \cite{rattazzi}, and $SU(5)
\times U(1)$ \cite{brahmhall,rattazzi}. They have been constructed such that
the only set of low-energy operators is $LL\Ebar$ or $LQ\Dbar$, or $\Ubar\Dbar
\Dbar$, respectively. There is thus no problem with proton decay. In order to
ensure that only the required set of non-renormalisable operators are allowed,
additional symmetries are required beyond the GUT gauge group. This is true for
both \rp-conservation and $\rpv$. Thus from a grand unified point of view there
is no preference for either \rp-conservation or $\rpv$.

{\it String Theory:}\ \ 
In string theories unification can be achieved without a simple gauge group.
There is thus no difficulty in having distinct quantum numbers for quarks and
lepton superfields. Indeed \rp-conserving and $\rpv$ string theories have been 
constructed \cite{bento}. At present, there does not seem to be a preference at
the string level for either of the two. 

In both string theory and in GUTs, there is no generic prediction for the size
of the $\rpv$-Yukawa couplings. This is analogous to the fermion mass problem.

{\it Discrete Gauge Symmetries:}\ \ There has been a further attack on this 
problem from a slightly different angle. If a discrete symmetry is a remnant of
a broken gauge symmetry it is called a discrete gauge symmetry. It has been 
argued that quantum gravity effects maximally violate all discrete symmetries
unless they are discrete gauge symmetries \cite{discrete1}. The condition that
the underlying gauge symmetry be anomaly-free can be translated into conditions
on the discrete symmetry. A systematic analysis of all ${\cal Z}_N$ symmetries
\cite{discrete2} has been performed. The result was that only two symmetries
were discrete gauge anomaly-free: \rp\ and baryon-parity \eq{bp}. Baryon-parity
was slightly favoured since in addition it prohibited dimension-5 proton-decay
operators. It has since been shown \cite{discrete3} that the non-linear 
constraints in \cite{discrete2} are model dependent thus possibly allowing an
even larger set of discrete symmetries.

Given the quantum gravity argument it is more appealing to determine the 
low-energy structure directly from gauge symmetries instead of discrete
symmetries. This can possibly even be connected with the fermion-mass
or flavour problem. This is an on-going field of research and it is too early 
to draw any conclusions. I just point out that gauged models with $\rpv$ have
been constructed \cite{gaugedrus,yuval}.

In conclusion, from the theoretical understanding of unification, there is no
clear preference between \rp\  and $\rpv$. In light of the very distinct 
phenomenology which we  discuss below, it is thus mandatory to experimentally 
search for both possibilities. \rp-conservation and $\rpv$ have the same 
minimal particle content. They also in principle have the same kind of
symmetries, as we have just argued: $G_{SM}$ plus an additional symmetry to 
protect the proton. They should thus both be considered as different versions 
of the MSSM. We shall denote the \rp\ conserving version of the MSSM as \rpmssm
and the \rp-violating version as \rpvmssm.

\section{Indirect Bounds}
The $\rpv$ interactions can contribute to various (low-energy) processes through
the virtual exchange of supersymmetric particles \cite{bhatt}. To date, all data
are in good agreement with the SM. This leads directly to bounds on the $\rpv$
operators. When determining such limits one must make some simplifying 
assumptions due to the large number of operators. In the following, we shall 
assume that one $\rpv$ operator at a time is dominant while the others are 
negligible. We thus do not include the sometimes very strict bounds on products
of operators, for example from $\mu\ra e\gamma$ \cite{beatrice}. This is an
important assumption but not unreasonable. It holds for the SM for example, 
where the top quark Yukawa coupling is almost a factor 40 larger than the
bottom Yukawa coupling. Since we do not know the origins of Yukawa couplings, we
do not know whether this is a generic feature. 

\begin{table}[t]
%\vspace{0.4cm}
\begin{center}
\begin{tabular}{|cc||cc|cc|cc||cc|}\hline
$ijk$ & $\lam_{ijk}$ & $ijk$ & $\lam'_{ijk}$ & $ijk$ & $\lam'_{ijk}$ &
$ijk$ & $\lam'_{ijk}$ & $ijk$ & $\lam''_{ijk}$ \\ \hline
121 & $0.05^{a\dagger}$ & 111 & $0.001^d$& 211 & $0.09^h$ & 311& $0.16^k$ 
& 112 & $10^{-6,\ell}$
 \\
122 & $0.05^{a\dagger}$ & 112 & $0.02^{a\dagger}$ & 212 & $0.09^h$ & 312& 
$0.16^k$ & 113& $10^{-5,m}$
 \\
123 & $0.05^{a\dagger}$ & 113 & $0.02^{a\dagger}$ & 213 & $0.09^h$ & 313
& $0.16^k$ & 123 & $1.25^{**}$ \\
131 & $0.06^b$  & 121 & $0.035^{e\dagger}$ & 221 & $0.18^i$ & 321& $0.20^{f*}$ 
& 212 & $1.25^{**}$   \\
132 & $0.06^b$  & 122 & $0.06^c$ & 222 & $0.18^i$ & 322& $0.20^{f* }$ 
& 213 &  $1.25^{**}$ \\
133 & $0.004^c$  & 123 & $0.20^{f*}$ & 223 & $0.18^i$ & 323& $0.20^{f*}$  & 
223 & $1.25^{**}$ \\
231 & $0.06^b$ & 131 & $0.035^{e\dagger}$ & 231 & $0.22^{j\dagger}$ & 331& 
$0.26^g$ & 
312 & $0.43^g$ \\
232 & $0.06^b$ & 132 & $0.33^g$ & 232 & $0.39^g$ & 332& $0.26^g$ & 313 & 
$0.43^g$ \\
233 & $0.06^b$ & 133 & $0.002^c$& 233 & $0.39^g$ & 333& $0.26^g$ & 323 & 
$0.43^g$ \\
\hline
\end{tabular}
\end{center}
\caption{Strictest bounds on $\rpv$ Yukawa couplings for ${\tilde m}=100\gev$.
The physical processes from which they are obtained are summarized in the main
text.\label{tab:bounds}} 
\end{table}

Before presenting the complete bounds, I shall discuss one example \cite{bgh} to
show how such bounds can be obtained. The operator $L_1L_3\Ebar_k$ can 
contribute to the decay $\tau\ra e\nu{\bar\nu}$ via the diagram in Figure 1b.
For large slepton masses, ${\tilde m}({\tilde e}^k_R)$, this interaction is
described by an effective 4-fermion Lagrangian (after Fierz re-ordering) 
\cite{bgh}
\beq
{\cal L}_{eff}= \frac{|\lam_{13k}|^2}{2{\tilde m}^2}
({\bar e}_L\gamma^\mu\nu_{eL})({\bar\nu}_{\tau L}\gamma_\mu\tau_L).
\eeq
This has the same structure as the term in the effective SM Lagrangian
and thus leads to an apparent shift in the Fermi constant for tau decays.
Considering the ratio $R_\tau\equiv\Gamma(\tau\ra e\nu{\bar\nu})/\Gamma
(\tau\ra\mu\nu{\bar\nu})$, the contribution from $\rpv$  relative to the
SM contribution is \cite{bgh}
\beq
R_\tau=R_\tau(SM)\left[1+2\frac{M_W^2}{g^2}\left(\frac{|\lam_{13k}|^2}{{\tilde
m}^2({\tilde e}^k_R)}\right)\right].
\eeq
Using the experimental value \cite{aleph} $R_\tau/R_\tau(SM)=
1.0006\pm0.0103$ we obtain the bounds
\beq
|\lam_{13k}| < 0.06\, \left(\frac{{\tilde
m}({\tilde e}^k_R)}{100\gev}\right),\quad k=1,2,3, 
\eeq
which are given in
Table 1. The strictest bounds on the remaining operators are 
also summarized in Table 1. 

The bounds in Table \ref{tab:bounds} are obtained from the following physical
processes: $^{(a)}$ charged current universality \cite{bgh,pdg}, $^{(b)}$ $
\Gamma(\tau\ra e\nu{\bar\nu})/\Gamma(\tau\ra\mu\nu{\bar\nu})$ \cite{bgh,pdg}, $^
{(c)}$ bound on the mass of $\nu_e$ \cite{hallsuzuki,tata,wark}, $^{(d)}$
neutrinoless double-beta decay \cite{moha,klapdor}, $^{(e)}$ atomic parity
violation \cite{davidson,wood,atomic}, $^{(f)}$ $D^0-{\bar D}^0$ mixing
\cite{wittig,gupta,agashe}, $^{(g) }$ $R_\ell =\Gamma_{had}(Z^0)/\Gamma_\ell (
Z^0)$ \cite{srid2,bhatt}, $^{( h)}$ $\Gamma(\pi\ra e{\bar\nu})/\Gamma(\pi\ra\mu
{\bar\nu})$ \cite{bgh}, $^{(i) }$ $BR(D^+\ra{\bar K}^{0*}\mu^+\nu_\mu)/BR(D^+
\ra {\bar K}^{ 0*}e^+\nu_\mu)$ \cite{bhattchoud,bhatt}, $^{(j)}$ $\nu_\mu$
deep-inelastic scattering \cite{bgh}, $^{(k)}$ $BR(\tau\ra\pi\nu_\tau)$
\cite{bhattchoud,bhatt}, $^{(\ell)}$ heavy nucleon decay \cite{sher}, and $^{(
m)}$ $n-{\bar n}$ oscillations \cite{zwirner,sher}.

The bounds are all given for ${\tilde m}=100\gev$ and they become weaker with
increasing ${\tilde m}$. They each depend on a specific scalar mass and have
various functional dependences on this mass.
\begin{equation}
\begin{array}{|l|c||c|c|}
\hline 
\lam_{ijk}& {m_{{\tilde e}_{Rk}}}/{100\gev} 
& \lam'_{111} & ({m_{{\tilde q}}}/{100\gev})^2 
({m_{{\tilde g}}}/{1\,TeV})^ {1/2} \\
\lam'_{11k},\lam'_{21k} & {m_{{\tilde d}_{Rk}}}/{100\gev} &
\lam'_{1j1}& {m_{{\tilde q}_{Lj}}}/{100\gev} \\
\lam_{133}, \lam'_{1jj} & \sqrt{ {m_{{\tilde\tau},{\tilde d}_j}}/{100\gev}} &
\lam'_{123} & \sqrt{ {m_{{\tilde b}_{R}}}/{100\gev}} \\
\lam'_{231} & { {m_{{\tilde \nu}_{\tau L}}}/{100\gev}} &
\lam'_{32k} & \sqrt{ {m_{{\tilde d}_{Rk}}}/{100\gev}}\\ \hline
\end{array}
\lab{scales}
\end{equation}
For $\lam'_{111}$ the dependence can be on either $m_{{\tilde d}_{Rk}}$, or $m_{
{\tilde u}_{Lk}}$. The bound in Table 1 is given for ${\tilde m}({\tilde g})=1\,
TeV$. For $\lam'_{132},$ $\lam'_{22k},$ $\lam'_{23k},$ $\lam'_{31k}$, and $\lam'
_{32k}$ one must consult the appropriate references since the dependence is only
given numerically. The bounds on $\lam''_{112,113}$ from heavy nucleon decay and
$n-{\bar n}$ oscillations have very strong mass dependences \cite{sher}. 

I have updated the previous bounds from charged current universality
\cite{bgh}, from lepton-universality \cite{bgh} and from \footnote{I thank
Gautam Bhattacharyya for providing me with updates on the bounds resulting from
$R_\ell $.} $R_\ell$ using more recent data \cite{pdg,aleph}. For the bound from
the electron neutrino mass I have used the upper bound \cite{wark} $m_{\nu_e}<5\
,eV$. The PDG number is $10-15\,eV$ \cite{pdg} and is very conservative 
\cite{lobashev}. The bound on $\lam$ from $m_{\nu_e}$ scales with the square 
root of the upper bound on $m_{\nu_e}$. For the bound from atomic parity
violation I have used the theory value \cite{atomic}: $Q_W^{th}=-73.17\pm0.13$.
The error includes the variations due to the unknown Higgs mass. I have also
used the recent new experimental number \cite{wood}. For the bound from
$D^0-{\bar D}^0$ mixing, I have updated the bound from \cite{agashe} to include
a lattice calculation of \cite{gupta} $B_D$ and a more updated value of $f_D$
\cite{wittig}. I have also included a $10\%$ error to account for the quenched
approximation. The bounds denoted by $^\dagger$ are $2\sigma$ bounds, the other
bounds are at the 1 sigma level. The bounds denoted by $^{(**)}$ are not direct
experimental bounds. They are obtained \cite{probir,sher} from the requirement
that the $\rpv$-coupling remains within the unitarity bound up to the grand
unified scale of $10^{16}\gev $. This need not be the case.

The bounds denoted by $^{*}$ are based on a further {\it assumption} about the
absolute mixing in the (SM) quark sector. As stated before, we do not know the
physical origin of Yukawa couplings or superpotential terms. It is a reasonable
(but not necessary) assumption that their structure is determined by some
symmetry at an energy scale well above the electroweak scale, \eg the GUT or
the Planck scale. We would expect this symmetry to be in terms of the weak {\it
current} eigenstates. Such a symmetry could then give us a single dominant
operator, for example 
\beq
L_1Q_1\Dbar_1=\lam'_{111}(-{\tilde e}_Lu_L{\bar d}_R+{\tilde\nu}_{eL}d_L
{\bar d}_R\ldots). 
\lab{rot}
\eeq
Below the electroweak scale the quarks become massive and we must rotate them
to their {\it mass} eigenstate basis. (The squarks must separately also be
rotated by a different rotation but that is not relevant to these bounds.) In 
\eq{rot} there are then separate rotations: $d_L\ra {\cal D}_{1j}d'_{jL}$ and 
$u_L\ra {\cal U}_{1j}u'_{jL}$ which generate extra $\rpv$ terms suppressed
by mixing angles \cite{rosshadro}.

For the quarks we do not know the {\it absolute} mixing of the down-quark 
sector, ${\cal D}_{ij}$, or of the up-quark sector ${\cal U}_{ij}$ and thus do
not know by how much to rotate the up- and down-quark current eigenstates. The {
\it relative} mixing of these two sectors is given by the CKM-matrix \cite{pdg} 
of the SM. If we assume the relative rotation is solely due to an absolute 
mixing in the up-quark sector (${\cal D}_{ij}={\bf 1}$) the best bounds are 
those given in Table 1. Those denoted by $^{*}$ are specifically based on this
mixing assumption. If however the relative mixing is solely due to absolute
mixing in the down-quark sector (${\cal U}_{ij}={\bf 1} $) the $D^0$-${\bar D}^0
$ mixing bounds no longer apply. There are then significantly stricter bounds on
many couplings from measurements of $K^+\ra\pi^ +\nu\nu$ decays \cite{agashe} 
\beq
\lam'_{ijk} < 0.012,\; (90\% CL),\quad j\not=3.
\lab{kpinunu}
\eeq
For Table 1 we have adopted the conservative estimate that the mixing is solely
due to the up-quark sector since we do not know the absolute mixing. We
therefore did not include the bounds \eq{kpinunu}.\footnote{There is a possible
loop-hole. The symmetry at the high energy scale could just produce such a
combination of couplings that is rotated to one single dominant coupling at low
energy. After all, it is possibly the same symmetry which produces the single
dominant quark Yukawa coupling in the SM. However, I do not adopt this
philosophy here.} 

\section{Changes to \rp-MSSM}
On the Lagrangian level the only change to the \rp-MSSM is the inclusion of the
operators in $W_\rpvm$ which give new lepton- and baryon number violating
Yukawa couplings. There are several changes in the phenomenology of
supersymmetry due to these couplings \cite{rosshadro}. 
\begin{itemize}
\begin{enumerate}
\item Lepton- or baryon number is violated as discussed in Sect. 5.
\item The LSP is not stable and can decay in the detector. It is no
longer a dark matter candidate.
\item The neutralino is not necessarily the LSP.
\item The single production of supersymmetric particles is possible.
\end{enumerate}
\end{itemize}

{\bf 2.} If for example the neutralino is the LSP and the dominant $\rpv$ 
operator is $L_1Q_2\Dbar_1$ it can decay as shown in Figure 1c. For LSP$={\tilde
\gamma}$ the decay rate is \cite{dawson,neutr} 
\beq
\Gamma_{{\tilde\gamma}}=\frac{3\alpha\lambda^{'2}_{121}}{128\pi^2} 
\frac{M^5_{\chi^0_1}}{{\tilde m}^4}.
\lab{gamlsp}
\eeq
The decay occurs in the detector if $c\gamma_L\tau({\tilde\gamma})\;
\lsim\; 1\,m$, or
\beq
\lam'_{121} > 1.4\cdot10^{-6}\sqrt{\gamma_L}
\left(\frac{{\tilde m}}{200\gev}\right)^2
\left(\frac{100\gev}{M_{\tilde\gamma}} \right)^{5/2}.
\lab{decaylength}
\eeq
where $\gamma_L$ is the Lorentz boost factor. This is well below the bound of 
Table 1. Recall also for comparison, that in the SM Yukawa couplings can be very
small: for the electron $h^e=3\cdot 10^{-6}$. We have presented these numerical
results for a photino for simplicity and clarity. The full analysis with a
neutralino LSP has been performed in \cite{peterhera,perez}. It involves
several subtleties due to the \rpmssm\ parameter space which can have
significant effects on the lifetime. Due to the LSP decay, supersymmetry with
broken \rp\  has no natural dark matter candidate.

{\bf 3.} In the \rpmssm the stable LSP must be charge and colour neutral for
cosmological reasons (\cf Chapter 15). In the \rpvmssm\ there is no preference 
for the nature of the unstable LSP. It can be any of the following \footnote{The
stop is listed separately since it has a special theoretical motivation
\cite{ellisrudaz}  and leads to quite distinct phenomenology given that the top
quark is so heavy.} 
\beq
LSP\quad\epsilon\quad
\{\chi^0_1,\chi^\pm_1,{\tilde g},{\tilde q},{\tilde t},
{\tilde \ell},{\tilde \nu} \}.
\eeq
In each case the collider phenomenology can be quite distinct.
 
{\bf 4.} In the \rpvmssm\  there are resonant and non-resonant single
particle production mechanisms. The resonant production mechanisms are
\barr
e^++e^-&\ra\; {\tilde \nu}_{Lj},\quad & L_1L_j\Ebar_1,\lab{res1} \\
e^-+u_j&\ra\; {\tilde d}_{Rk},\quad & L_1Q_j\Dbar_k, \lab{res2}\\
e^-+{\bar d}_k&\ra\; {\tilde {\bar u}}_{Lj},\quad &L_1 Q_j\Dbar_k, \\
{\bar u}_j+{d}_k &\ra\; {\tilde e}^-_{Li}, \quad & L_iQ_j\Dbar_k, \\
d_j+{\bar d}_k&\ra\;{\tilde{\bar\nu}}_{Li},\quad& L_1Q_j\Dbar_k, \\
{\bar u}_i+{\bar d}_j &\ra\; {\tilde {d}}_{Rk}, \quad & \Ubar_i\Dbar_j\Dbar_k,\\
d_j+d_k &\ra\; {\tilde {\bar u}}_{Ri},\quad& \Ubar_i\Dbar_j\Dbar_k \lab{res4}
\earr
These processes can be realized at $e^+e^-$-colliders, at HERA, and at hadron
colliders, respectively. There are many further t-channel single sparticle
production processes. For example at an $e^+e^-$-collider, we can have $e^++e^-
\ra {\tilde \chi}^0_1+\nu_j$ via t-channel selectron exchange. The t-channel
exchange of squarks (sleptons) can also contribute to $q{\bar q}$ ($\ell{\tilde
\ell}$) pair production, leading to indirect bounds \cite{opal}.

\section{Collider Phenomenology}
The supersymmetric signals for $\rpv$ will be a combination of supersymmetric 
production and decay to \rp\ even final states. Supersymmetric particles can 
be produced in pairs via MSSM gauge couplings or singly as in 
\eq{res1}-\eq{res4}. The former benefits from large couplings while being 
kinematically restricted to masses $<\sqrt{s}/2$. The latter case has double 
the kinematic reach but suffers from typically small Yukawa couplings. 
Combining the various production modes with the decays and the different 
dominant operators leads to a wide range of potential signals to search for. 
Instead of systematically listing them I shall focus on two examples. 
Throughout we shall assume a neutralino LSP. 

\subsection{Squark Pair Production at the Tevatron}
Squark pair production at the Tevatron proceeds via the known gauge couplings
of the $R_p$-MSSM
\beq
q{\bar q},gg\ra {\tilde q} + {\bar{\tilde q}}.
\eeq
In $\rpv$, once produced the squarks decay to an \rp\  even final state. 
Let us consider a dominant $L_iL_j\Ebar_k$
operator. The couplings $\lam_{ijk}$ are bounded to be smaller than gauge 
couplings. Thus we expect the squarks to cascade decay to LSPs as in the MSSM. 
The LSPs in turn will then decay via the operator $L_iL_j\Ebar_k$ to two 
charged leptons and a neutrino each (\cf Figure 1c). If each squark decays 
directly to the LSP (assuming it is the second lightest)
\beq
q'{\bar q}',gg\ra {\tilde q} + {\bar{\tilde q}}\ra q{\bar q}+{\tilde\chi}^0_1
{\tilde\chi}^0_1 \ra q{\bar q}+ l^+l^-l^+l^-\nu\nu.
\lab{squarkpair}
\eeq
\begin{table}[t]
\begin{center}
\begin{tabular}{|c|c|c|c|c|c|}\hline
&$L_{1,2}Q_{1,2}\Dbar_k$, & $L_{1,2}L_3\Ebar_3$&
$L_{1}L_2\Ebar_3$ & $L_{1,2}L_3\Ebar_{1,2}$ & $L_1L_2\Ebar_{1,2}$ \\ \hline
$m_{\tilde q}$  &$100\gev$ &$100\gev$ &$140\gev$ & $160\gev$ &$175\gev$\\\hline
\end{tabular}
\end{center}
\caption{Squark mass bounds from the Tevatron for various dominant
$\rpv$-operators $^{45}$.}
\vspace{-0.5cm}
\end{table}
We therefore have a multi-lepton signal which is detectable \cite{dp}. To date 
it has not been searched for with $\rpv$ in mind. However, before the top quark 
discovery there was a bound from CDF on a di-lepton production cross section. 
Making corresponding cuts and with some simple assumptions this can be
translated into a bound on the rate of the process \eq{squarkpair} and thus a 
lower bound on the squark mass \cite{dp}. The assumptions are: (i) $BR({\tilde
q}\ra{\tilde\gamma}q)=100\%$, ($m_{\tilde q}<m_{\tilde g}$), (ii) LSP$={\tilde
\gamma}$ with $M_{{\tilde\gamma}}=30\gev$, (iii) $\lam,\lam'$ satisfy the bound
 \eq{decaylength}. For various dominant operators the bounds are given in Table 
2. No attempt was made to consider final state $\tau$'s due to lack of data. 
These bounds are comparable to the \rpmssm squark mass bounds. Since, the
theoretical analysis has been improved to allow for neutralino LSPs, more
involved cascade decays and the operator $\Ubar\Dbar\Dbar$
\cite{rosshadro,tevatron}. However, to date no experimental analysis has been
performed.

\subsection{Resonant Squark Production at HERA}
HERA offers the possibility to test the operators $L_1Q_j\Dbar_k$ via resonant
squark production \cite{herasquark} \footnote{HERA has accumulated most of its
data as a positron  proton collider.}
\barr
e^++d_k&\ra& {\tilde u}_j \ra (e^++d_k,\;{\tilde\chi}^0_1+u_j,\;{\tilde\chi}^+_1
+d_j),\\
e^++{\bar u}_j&\ra& {\bar{\tilde d}}_k \ra (e^++{\bar u}_j,\; {\bar\nu}_e+
{\bar d}_j,\; {\tilde\chi}^0_1+ {\bar d}_k).
\earr
We have included what are most likely the dominant decay modes. The neutralino
and chargino will decay as in Figure 1c
\beq
{\tilde \chi}^0_1\ra (e^\pm,\,\nu)+{\rm 2\,jets},\quad {\tilde \chi}^+_1\ra
(e^+,\,\nu)+{\rm 2\,jets}. 
\eeq
The neutralino can decay to the electron or positron since it is a Majorana
fermion. We are thus left with several distinct decay topologies. {\it (i)} If 
the squark is the LSP it will decay to $e^++q$ or ${\bar \nu}_e +q$
(${\bar{\tilde d}}_k$). The first looks just like neutral current DIS, except
that for $x_{Bj} \approx {\tilde m}^2({\tilde q})/s$ it results in a flat
distribution in $y_e$ whereas NC-DIS gives a $1/y_e^2$ distribution. The latter
looks just like CC-DIS. {\it (ii)} If the gauginos are lighter than the
squark the gaugino decay will dominate \cite{perez} \footnote{The gaugino
decays could be suppressed by phase space or by partial cancellations of the
neutralino couplings \cite{peterhera,ellis}.}. The clearest signal is a
high $p_T$ electron which is essentially background free. The high $p_T$
positron or the missing $p_T$ of the neutrino can also be searched for. 

All five signals have been searched for by the H1 collaboration \cite{H1rpv} 
in the 1994 $e^+$ data (${\cal L}=2.83\,pb^{-1}$). The observations were in 
excellent agreement with the SM. The resulting bounds on the couplings are
summarized in Table 3.
\begin{table}[t]
\caption{Exclusion upper limits at $95\%$ CL on $\lam'_{1jk}$ for ${\tilde m}
({\tilde q})=150\gev$ and ${\tilde m}({\tilde\chi}^0_1)=40\gev$ for two 
different dominant admixtures of the neutralino.}
\begin{center}
\begin{tabular}{|c||c|c|c|c|c|c|c|c|c|} \hline
&$\lam'_{111}$ &$\lam'_{112}$ & $\lam'_{113}$ & $\lam'_{121}$ & $\lam'_{122}$ &
$\lam'_{123}$ & $\lam'_{131}$ & $\lam'_{132}$ & $\lam'_{133}$ \\ \hline \hline
${\tilde\gamma}$-like &0.056&0.14& 0.18 & 0.058 & 0.19 & 0.30&0.06&0.22&0.55 \\
\hline  
${\tilde Z}^0$-like &0.048 & 0.12 &0.15 &0.048 & 0.16 &0.26 &0.05 &0.19 &0.48 \\
\hline
\end{tabular}
\end{center}
\vspace{-0.5cm}
\end{table}
After rescaling the bounds of Table 1 we see that the direct search is an
improvement for $\lam'_{121},\,\lam'_{131},$ and $\lam'_{132}$. In the more 
recent data, an excess has been observed in high $Q^2$ NC-DIS \cite{data}. If 
this persists it can possibly be interpreted as the resonant production of a 
squark via an $L_1Q_j\Dbar_k$ operator \cite{ellis,highq2rpv}. 

\section{Cosmology}
\subsection{Bounds from GUT-Scale Baryogenesis}
There is a very strict bound on all $\rpv$ Yukawa couplings assuming the 
presently observed matter-asymmetry was created above the electroweak scale, \eg
at the GUT-scale \cite{gutbound} which I briefly recount. Assume that at the GUT
scale a baryon- and lepton-asymmetry was created with possibly both $B+L\not=0$
and $B-L\not=0$. The electroweak sector of the SM (and MSSM) has baryon-number
and lepton-number violating ``sphaleron'' interactions which conserve $B-L$ but
violate $B+L$. These are in equilibrium above the electroweak phase transition
and they thus erase the $B+L\not=0$ component of the matter asymmetry. 

Consider now adding one additional $\rpv$ operator, \eg $\Ubar\Dbar\Dbar$ which
violates baryon number. If it is in thermal equilibrium during an epoch after
the GUT epoch and together with the sphaleron-interactions then together they
will erase the entire matter asymmetry. In order to avoid this scenario the
$\rpv$ interactions should not be in thermal equilibrium above the electroweak
scale resulting in the bounds \cite{gutbound,bound2,grahamsphal} 
\beq
\lam,\lam',\lam'' <  5\cdot10^{-7} \left(\frac{{\tilde m}}
{1\,TeV}\right)^{1/2}.
\lab{gutbound}
\eeq
It should be clear that the argument holds for $LL\Ebar$ or $LQ\Dbar$ operators
as well if lepton flavour is universal \cite{gutbound}. This is an extremely
strict bound on {\it all} the couplings. If it is valid then $\rpv$ is
irrelevant for collider physics and can only have cosmological effects. 

There are two important loop-holes in this argument. The first and most obvious
one is that the matter genesis occurred at the electroweak scale or below
\cite{carena}. The second loop-hole has to do with the inclusion of all the
symmetries and conserved quantum numbers \cite{grahamsphal,nelson,cdeo3}. The
electroweak sphaleron interactions do not just conserve $B-L$. They conserve
the three quantum numbers $B/3-L_i$, one for each lepton flavour. These can
also be written as $B-L$ and two independent combinations of $L_i-L_j$. First,
again consider an additional $\Ubar\Dbar\Dbar$ operator. If the matter genesis
at the GUT scale is asymmetric in the lepton flavours, $(L_i-L_j)|_{M_{GUT}}\not
= 0$, then this lepton-asymmetry is untouched by the sphalerons and by the $
\Delta B \not=0$ operators $\Ubar\Dbar\Dbar$ operators. The baryon asymmetry is
however erased. Below the electroweak scale, the lepton-asymmetry is partially
converted into a baryon-asymmetry via (SM) leptonic {\it and} supersymmetric 
mass effects \cite{grahamsphal}. If now instead, we add a lepton-number
violating operator, we will retain a matter asymmetry as long as one lepton
flavour remains conserved. In order for $L_\tau$ for example to be conserved,
all $L_\tau$ violating operators must remain out of thermal equilibrium above
the elctroweak scale, \ie satisfy the bound \eq{gutbound}. From the low-energy
point of view this is completely consistent with our Ansatz of considering only
one large dominant coupling at a time. Thus in these simple scenarios the
bounds \eq{gutbound} are evaded. 

\subsection{Long-Lived LSP}
One can consider three distinct ranges for the lifetime of the LSP 
\beq
(i)\;\tau_{LSP}\,\lsim\, 10^{-8}s,\quad
(ii)\;10^{-8}s\,\lsim\,\tau_{LSP}\,\lsim\, 10^{7}\tau_{u},\quad
(iii)\;\tau_{LSP}\,>\,10^7\tau_{u},
\eeq
where $\tau_u\sim10^{10}yr$ is the present age of the universe. We have
discussed the first case in detail in the previous chapters. The third case is
indistinguishable from the \rpmssm with the LSP being a good dark matter
candidate.  In the second case, the LSP can provide a long-lived relic whose
decays can potentially lead to observable effects in the universe. There are
bounds excluding any such relic with lifetimes \cite{subir} 
\beq
1s<\tau_{LSP}<10^{17}yr.
\eeq
The lower end of the excluded region is due to the effects of hadron showers 
{}from LSP decays on the primordial abundances of light nuclei \cite{reno}. The 
upper bound is from searches for upward going muons in underground detectors
which can result from $\nu_ \mu$'s in LSP decays \cite{gelmini}. Note that even
if $\tau_{LSP}>\tau_u$ the relic abundance is so large \footnote{This is for
most values of the MSSM parameters, \cf Chapter 15.} that the decay of only a 
small fraction can lead to observable effects. 

The above restrictions on decay lifetimes can be immediately applied to the case
of \rpvmssm. If we include LSP decays in collider experiments we are left with
a gap of eight orders of magnitude in lifetimes $10^{-8}s <\tau_{LSP} < 1s$ 
where no observational tests are presently known. It is very important to find
physical effects which could help to close this gap. Since the lifetime depends
on the square of the $\rpv$ Yukawa coupling this corresponds to a gap of four 
orders of magnitude in the coupling. For a photino LSP we can translate the 
above bounds into bounds on the $\rpv$ Yukawa couplings \cite{bound2}. Using
Eq.\eq{gamlsp} we obtain the excluded region for the couplings
\beq
10^{-22}< (\lam,\lam',\lam'')\cdot \left(\frac{200\gev}{{\tilde m}}\right)^2
\left(\frac{M_{\tilde\gamma}}{100\gev}\right)^{5/2} <10^{-10}.
\eeq
Note that the lower range of these bounds extends well beyond the already strict
bound from proton decay \eq{proton}. For a generic neutralino LSP the lifetime
depends strongly on the MSSM parameters \cite{peterhera,perez} and the bounds
can only be transferred with caution.

\section{Outlook}
Once we include the $\rpv$ terms in the superpotential we are left with a 
bewildering set of possibilities. We have 45 new Yukawa couplings of which 
any could be dominant and we have a set of seven different potential LSPs, each
possibly leading to quite different phenomenology. This situation requires a 
systematic approach. 

I would here like to suggest a two-fold approach. The theoretically best
motivated model is one based on universal soft breaking terms at the
unification scale $\sim10^{16}\gev$, completely analogous to the MSSM. To
obtain the low-energy spectrum one then employs the renormalisation group
equations {\it including} the $\rpv$-Yukawa couplings and all the soft breaking
terms. This program has yet to be completed \cite{rpvrges,beatrice}. However,
since most of the $\rpv$-Yukawa couplings are bounded to be relatively small we
expect for large regions in parameter space the spectrum of the \rpvmssm\ to
look just like that of the \rpmssm. The only difference will be a decaying
neutralino LSP. To this extent the program has been implemented in SUSYGEN
\cite{susygen}, the supersymmetry Monte Carlo generator for $e^+e^-$-colliders.
$\rpv$ has only been implemented partially in ISAJET, a generator for hadron 
colliders \cite{isajet}.

As a second step, I suggest a systematic listing of potential signal topologies
which can arise for spectra not obtained in the simple unification approach.
Any exotic topologies can easily be searched for on a qualitative level. These
two approaches combined should ensure that we do not miss any signal for
supersymmetry and also do not end up searching vigorously in the $\rpv$-hat
every time an experimental anomaly appears.

\section*{Acknowledgments}
I would like to thank the many people I have collaborated with on R-parity
violation:  Ben Allanach, Jon Butterworth, A. Chamseddine, Manoranjan Gu\-chait,
Smaragda Lola, Peter Morawitz, Felicitas Pauss, Emmanuelle Perez, Roger
Phillips, Heath Pois,  D.P. Roy and Yves Sirois. In particular I would like to
thank Graham Ross who got me started on the subject and continuously encouraged
me. I would furthermore like to thank Sacha Davidson and Gautam Bhattacharyya
for discussions on updating the indirect bounds. I would like to thank Subir
Sarkar for very helpful discussions on the long-lived LSP. I thank Ben Allanach,
Gautam Bhattacharyya, Sacha Davidson, Gian Giudice, and Subir Sarkar for reading
the manuscript and their helpful comments.


\begin{thebibliography}{99}
\bibitem{neutr} The more general formula for a neutralino LSP can be found
in \cite{peterhera}.
\end{thebibliography}
\end{document}